\documentclass[11pt]{article}
 \usepackage{sao2}
 \usepackage{times}
 \usepackage{graphicx}
 \usepackage{psfig}
 \usepackage{amsmath}
 \usepackage{amssymb}
\begin{document}
 \title{$BVR_cI_c$ photometry of the GRB~980703 and GRB~990123
        host galaxies}
 \author{V. V. Sokolov, T. A. Fatkhullin, V. N. Komarova}

     \institute{Special Astrophysical Observatory of R.A.S.,
  Karachai-Cherkessia, Nizhnij Arkhyz, 357147 Russia; sokolov@sao.ru}



     \maketitle
     
\begin{abstract}
We present a photometry of GRB~980703 and GRB~990123 host galaxies which was
performed about 20 days and a half year after gamma-ray bursts occured,
respectively. The contributions of the optical transiets (OT) were
negligible in
both cases. We derived broad band $BVR_cI_c$ spectra of the host
galaxies and compared them to continuum spectra of different Hubble-type
galaxies and averaged spectra of starburst galaxies.
For $H_0=60\ km\ s^{-1}
Mpc^{-1}$ and three Friedmann cosmological models with matter density and
cosmological constant parameters ($\Omega_m$,$\Omega_\Lambda$) = (1,0),(0,0),
(0,1) we estimated $M_{B_{rest}}$ and star-forming rates (SFRs)
using the fluxes in photometric bands for the host galaxies.
Within the range of cosmological parameters our estimates of the absolute
magnitudes are: $M_{B_{rest}} = -20.60 \div -21.73$ for the GRB~980703 host
galaxy and $M_{B_{rest}} = -20.20 \div -21.82$ for the GRB~990123 host
galaxy.
We obtained estimates of K-correction values and absolute magnitudes 
of the host galaxies using SEDs for star-forming galaxies.


\end{abstract}

\section{Introduction}

The origin of cosmic $\gamma$-ray bursts (GRBs) is still one of
the outstanding problems in modern astronomy. Recently the study of GRBs
has been revolutionized by the discovery of X-ray (Costa et al. 1997),
optical (van Paradijs et al. 1997) and radio transients (Frail et al. 1997).
The follow-up of optical transients (OT) resulted in the majority
not only with OTs' light curves but also their redshift measurements  and
the detection of GRB hosts. 
By now about 15 optical afterglows have been detected, almost all within a
fraction of an arc second of very faint galaxies, with typical R-band
magnitudes of 22-26 and over. About 10 redshifts have been measured in the
range from $z = 0.6$ to $z = 3.4$. 
The long duration GRBs appear to be associated 
with star forming regions in the galaxies.
The studies of the host galaxies and their properties
advance understanding of the nature of the progenitor systems.
In some proposed models an association of bursts
with explosions of massive stars become popular (see ref. Paczyn\'ski, 1999 and
MacFadyen A. \& Woosley S. E., 1999 and references therein). \par
Here we report our direct $BVR_cI_c$ imagings of the GRB~980703 optical
counterpart which were obtained after observations of Bloom et al.~(1998).
The latest observations of this object were obtained in the infrared $JHK$ bands
by Bloom et al.~(1998) In the case of the GRB~990123 host galaxy our $BVR_cI_c$
observations are the latest.
In this paper we compare the $BVR_cI_c$ spectra of the GRB~980703 and 
GRB~990123 host galaxies with spectral energy distributions of normal 
galaxies of different Hubble types and extend the comparision to averaged
spectra (SEDs) of S1 and S2 star-forming galaxies (Connoly et al.,~1995).
We obtained estimates of K-correction values and absolute magnitudes 
of the host galaxies using spectra 
of normal Hubble-type galaxies and SEDs for star-forming galaxies. 
Rough estimates of low limits for star-forming rates (SFR) in the
GRB~980703 and GRB~990123 host galaxies were performed using the continuum
luminosity at $\lambda = 2800$\AA \ by extrapolating for the GRB~990123 host
galaxy between $R_c$ and $I_c$ bands and by value of the flux in the $V$ band for
the GRB~980703 host galaxy.

\section{Observations and data reduction}
Observations of the host galaxies of GRB~980703 and GRB~990123 were performed
using the primary focus CCD photometer of the 6m telescope of SAO RAS.
It was carried out with the standard (Johnson-Kron-Cousins) photometric
$BVR_cI_c$ system. The direct $BVR_cI_c$ images for the GRB 980703 host galaxy
are given in {\small \verb"http://www.sao.ru/~sokolov/GRB/980703.html"}
 Table~\ref{phot} presents the summary of observations.
\footnote{This paper gives more accurate $BVR_cI_c$ values with the UT dates
(corresponding to Table~\ref{phot}) of observations for the host galaxy of
GRB 980703 unlike the preliminary $BVR_cI_c$ values given in
GCN notice \#147.}.
\begin{table*}
\caption{Photometry of the host galaxies of the GRB~980703 and GRB~990123}
\label{phot}
\begin{center}
\begin{tabular}{lllccc}
\hline
\hline
Host       & Date UT         & Band  & Exp. & Magnitude, obs  & Seeing    \\
	   &                 &       & (s)  &                 &           \\
\hline
GRB~980703 & 24.05 Jul. 1998 & $B$   & 480  & $23.40\pm 0.12$ & $1\arcsec.3$ \\
	   & 24.06 Jul.      & $V$   & 320  & $22.85\pm 0.10$ & $1\arcsec.2$ \\
	   & 24.06 Jul.      & $R_c$ & 300  & $22.44\pm 0.08$ & $1\arcsec.2$ \\
	   & 24.07 Jul.      & $I_c$ & 360  & $22.26\pm 0.18$ & $1\arcsec.2$ \\
	   &                 &       &      &                 &           \\
GRB~990123 & 8.85 Jul.       & $B$   & 600  & $24.90\pm 0.16$ & $1\arcsec.5$ \\
	   & 8.86 Jul.       & $V$   & 600  & $24.47\pm 0.13$ & $1\arcsec.3$ \\
           & 8.84 Jul. 1999  & $R_c$ & 600  & $24.47\pm 0.14$ & $1\arcsec.1$ \\
           & 8.87 Jul.       & $I_c$ & 600  & $24.06\pm 0.3$  & $1\arcsec.3$ \\
\hline
\hline
\end{tabular}
\end{center}
\end{table*}
We performed photometric calibrations using the Landolt standard fields
(Landolt,~1992):
PG1633, PG1657, PG2331, PG2336 for the host galaxy of GRB980703, and
PG2213, SA110 for the host galaxy of GRB990123.
To estimate the Milky Way redenning for the host galaxy of  GRB980703,
we used extinction values from Cardelli, Clayton \& Mathis (1989).
We derived $0^{m}\!.251$, $0^{m}\!.188$, $0^{m}\!.141$ and $0^{m}\!.090$  in our
photometric bands $BVR_cI_c$ respectively. Our direct $BVR_cI_c$ imagings
of the GRB~980703 optical counterpart were obtained after observations of
Bloom et al.~(1998) on July 18 UT.
The most recent observations of this object were obtained in the infrared
$JHK$ bands by Bloom et al.~(1998) on August 7 UT.
In the case of the GRB~990123 host galaxy our $BVR_cI_c$ observations are the
latest. We consider that the continuum was expected with minimum change
due to the fading of the OT.

\section{Cosmological models}
   The estimate of intrinsic physical parameters of extragalactic objects
  with redshifts approaching $1$ depends considerably on the
  adopted cosmological model.  The standard Friedmann model contains
    three parameters: Hubble constant $H_0$, matter density parameter
    $\Omega_m$, and cosmological constant parameter $\Omega_\Lambda$.
    Recent studies of the Hubble constant put it within the range 50 -- 70
    km s$^{-1}$ Mpc$^{-1}$ (see Theureau et al.~1997).
   In this paper we adopt $H_0 = 60$ km s$^{-1}$ Mpc$^{-1}$ .
   The values of $\Omega_m$ and $\Omega_\Lambda$ are
    observationally less constrained.  The recent work on the $m-z$
    test with supernovae of
    type Ia at redshifts up to $1$ by Garnavich et al.~(1998) make it
    imperative to consider in addition to the standard inflationary model
    also an empty universe with a cosmological constant.
    For a review of modern cosmological models and the necessary mathematical
    relations, see Baryshev et al.~(1994). Here we use  three Friedmann
    models which conveniently limit reasonable possibilities:
 \[ H_0 = 60 \mbox{ km s$^{-1}$ Mpc$^{-1}$, } \Omega_m = 1\mbox{, }\Omega_\Lambda
 = 0 \mbox{~~  (A)} \]
 \[ H_0 = 60 \mbox{ km s$^{-1}$ Mpc$^{-1}$, } \Omega_m = 0\mbox{, }\Omega_\Lambda
 = 0 \mbox{~~  (B)} \]
 \[ H_0 = 60 \mbox{ km s$^{-1}$ Mpc$^{-1}$, } \Omega_m = 0\mbox{, }\Omega_\Lambda
 = 1 \mbox{~~  (C)} \]

 For these models the relation $\Omega_m + \Omega_\Lambda
    + \Omega_k
 = 1$ is valid, where $\Omega_m = \rho_0 8 \pi G/3H_0^2$, $\Omega_\Lambda=
   \Lambda c^2/3H_0^2$, and $\Omega_k = -kc^2/R_0^2H_0^2$.  Here $\rho$,
 $\Lambda$, $k$, and $R$ are density, cosmological constant, curvature constant,
 and radius of curvature, respectively, and ``0" denotes the present epoch.

  The luminosity distance $R_{lum}$, the angular size distance $R_{ang}$
  and the proper metric distance $R_{p}$ are connected by the relation:
 \begin{equation}
 \label{Rlum}
 R_{lum} = R_{ang}(1+z)^2 = R_p(1+z)
 \end{equation}
 where the proper distances for the adopted models are given by
 \begin{equation}
 R_p=
 \begin{cases}
 R_H\: \frac{2(z-\sqrt{1+z}+1)}{1+z} & \text{ for model A,} \\
 R_H\: \frac{z(1+0.5z)}{1+z} & \text{  for model B,} \\
 R_H\: z& \text{ for model C.} \\
 \end{cases}
 \end{equation}
   Here $R_H = c/H_0$ is the present value of the Hubble radius. \par
  The absolute magnitude $M_{(i)}$ of the source observed in filter ($i$) can
  be calculated from the magnitude-redshift relation
 \begin{equation}
 \label{absmagn}
 M_{(i)} = m_{(i)} - K_{(i)}(z) -5\log(R_{lum}/{\rm Mpc}) - 25
 \end{equation}
  where $m_{(i)}$ is the observed magnitude of the object in the photometric
  band system ($i$) and $K_{(i)}(z)$ is the K-correction at redshift $z$,
    calculated from the rest-frame spectral energy distribution.

    The linear size $l$ of an object having an angular size $\theta$ is given
 by
   \begin{equation}
    \label{linsize}
    l = \theta R_{ang} = \theta R_p /(1+z)
   \end{equation}
 The K-correction in Eq. (3) can be calculated from the standard formula
   (Oke \& Sandage 1968):
 \begin{eqnarray}
  \label{k-corr}
  K_{(i)}(z) & = & 2.5\log(1+z)+ \nonumber \\
  & + & 2.5\log\frac{\int_0^\infty F_{\lambda} S_{(i)}(\lambda) {\rm d}\lambda}
 {\int_0^\infty F_{\lambda/(1+z)} S_{(i)}(\lambda) {\rm d}\lambda}
 \end{eqnarray}
 In this formula $F_{\lambda}$ is the rest-frame spectral energy
 distribution, $S_{(i)}$ is the sensitivity function for the filter (i).
 For our photometric system we used sensitivity functions for $BVR_cI_c$
 filters from  Bessel M. (1990).

\section{Estimation of the K-corrections}
The first spectral observations of the OT of  GRB~980703 were obtained
with the Keck-II 10-m telescope on UT 1998 July 07.6 and 19.6
(Djorgovski et al.~1998). Several strong emission lines were detected.
These were [OII], H$\delta$, H$\gamma$, H$\beta$ and [OIII] with redshift
$z=0.9662 \pm 0.0002$. In addition, in the blue part of the spectrum some
absorbtion features (FeII and MgII, MgI - absorbtion systems) with
$z=0.9656\pm 0.0006$ were detected. 
In the case of the OT of GRB~990123 absorbtion lines were detected only
(Kelson et al.~1999, Hjorth et al.~1999) with $z_{abs}=1.6004$.
The HST image reveals that the optical transient is offset by $0\arcsec.67$
from an extended object (Bloom et al.~1999). This galaxy is most likely to be
a host galaxy of GRB~990123 and source of the absorbtion lines of metals
at redshift $z=1.6004$. \par
\begin{figure*}[t]
\begin{center}
\includegraphics[height=9.0cm,angle=-90,bb=85 5 585 722,clip]{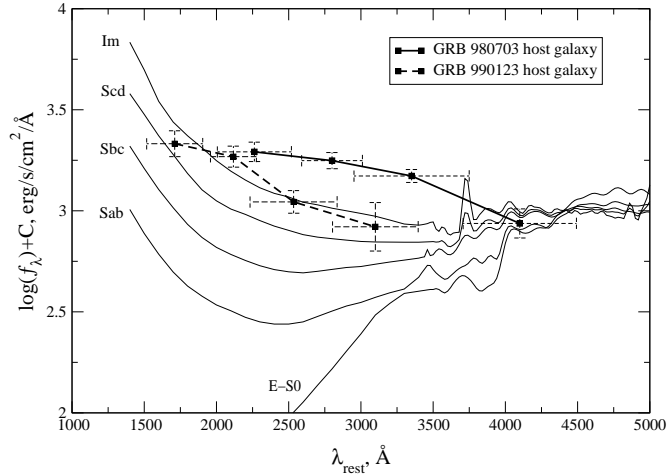}
\caption{A comparision of the GRB~980703 and GRB~990123 host galaxy broad
         band rest-frame
         ($z=0.966$ and $z=1.6004$ respectively)
	 spectra $\log F_{\lambda}=\log F_{\lambda, obs}+C$ to averaged
	 continuum spectra of galaxies of different Hubble types. The spectra
	 were shifted by some arbitrary constants for the best fits.
	 FWHM of each filter for $\lambda_{eff}$ with the account for $z$
	 are denoted by dashed horizontal lines with bars.}
\label{GRB_HT}
\end{center}
\end{figure*}

In Figure~\ref{GRB_HT} we compared the $BVR_cI_c$ broad band spectra of
the host galaxies of GRB~980703 and GRB~990123 with the spectral energy
distributions of normal galaxies of different Hubble types (Pence, 1976).
Table~\ref{fluxes} presents fluxes of the host galaxies. Here we used
the zero-points from Fukugita et al. (1995) for our photometric bands.
For the GRB~980703 the host galaxy fluxes are presented according to dereddened
magnitudes. For the GRB~990123 within our magnitude errors the Galactic
reddening is negligible.
\begin{table*}
\caption{Fluxes of the host galaxies}
\label{fluxes}
\begin{center}
\begin{tabular}{lccc}
\hline
\hline
Host       & Band &  $\log Flux_{\lambda, obs}$     &     $\log Flux_{\nu, obs}$       \\
           &      & $\frac{erg}{s \cdot cm^2 \mbox{\AA}}$  & $\frac{erg}{s \cdot cm^2 Hz}$ \\
\hline
GRB~980703 &  $B$ & $-17.468\pm 0.048$              & $-28.656\pm 0.048$        \\
           &  $V$ & $-17.508\pm 0.040$              & $-28.509\pm 0.040$       \\
	   & $R_c$& $-17.588\pm 0.032$              & $-28.440\pm 0.032$       \\
	   & $I_c$& $-17.823\pm 0.072$              & $-28.483\pm 0.072$       \\
           &      &                                 &                      \\
GRB~990123 & $B$  & $-18.168\pm 0.064$	            & $-29.356\pm 0.064$       \\
           & $V$  & $-18.231\pm 0.052$              & $-29.233\pm 0.052$       \\
	   & $R_c$& $-18.456\pm 0.056$              & $-29.308\pm 0.056$       \\
	   & $I_c$& $-18.579\pm 0.120$              & $-29.247\pm 0.120$       \\	   
\hline
\hline
\end{tabular}
\end{center}
\end{table*}
The central $\lambda_{obs}=\lambda_{eff}$ for our photometric system are
equal correspondingly to: $\lambda_B=4448\mbox{\AA}$, $\lambda_V=5505\mbox{\AA}$,
$\lambda_R=6588\mbox{\AA}$ and $\lambda_I=8060\mbox{\AA}$, FWHM are equal to:
$\Delta\lambda_B=1008\mbox{\AA}$, $\Delta\lambda_V=827\mbox{\AA}$,
$\Delta\lambda_R=1568\mbox{\AA}$,
$\Delta\lambda_I=1542\mbox{\AA}$, respectively.\par
Using equation \ref{k-corr} and the spectral energy distribution of the
Im Hubble-type galaxies 
we estimated the value of the K-correction for the magnitudes 
of the host galaxies of GRB~980703 and GRB~990123 according to $z=0.966$ and
$z=1.6$, respectively. The estimated values of the K-correction in the
$B$-band are $K_B=0.68$ for the GRB~980703 host galaxy and $K_B=0.88$ for the
GRB~990123 host galaxy.
In this case, the absolute magnitudes
from equation \ref{absmagn} for the host galaxy of GRB~980703 are:
$M_{B_{rest}}=-21.29$ for model~(A),
$M_{B_{rest}}=-21.81$ for model~(B) and
$M_{B_{rest}}=-22.42$ for model~(C).
For the host galaxy of GRB~990123 in the same way
absolute magnitudes are:
$M_{B_{rest}}=-20.95$ for model~(A),
$M_{B_{rest}}=-21.77$ for model~(B) and
$M_{B_{rest}}=-22.57$ for model~(C). \par
\begin{figure*}[t]
\begin{center}
\includegraphics[height=9.0cm,angle=-90,bb=85 5 585 722,clip]{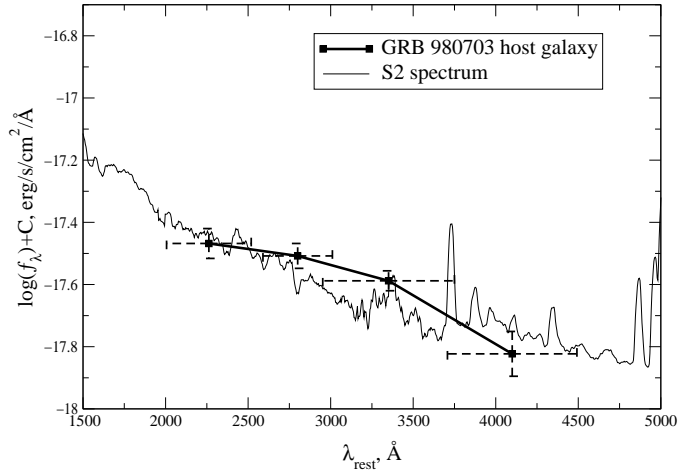}
\caption{A comparision of the GRB~980703 host galaxy broad
         band rest-frame ($z=0.966$)
	 spectrum  to spectrum of S2-galaxies $\log F_{\lambda}=\log F_{\lambda, S2}+C$
	 (see Connoly et al.,~1995). 
	 FWHM of each filter for $\lambda_{eff}$ with the account for $z$
	 are denoted by dashed horizontal lines with bars.}
\label{grb980703_S2}
\end{center}
\end{figure*}
\begin{figure*}[t]
\begin{center}
\includegraphics[height=9.0cm,angle=-90,bb=85 5 585 722,clip]{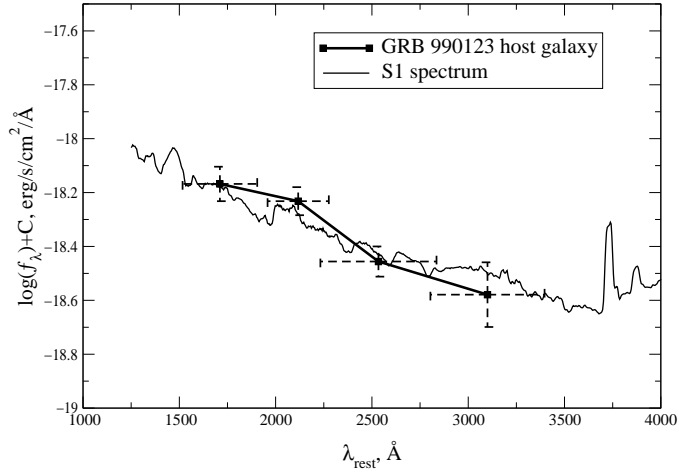}
\caption{A comparision of the GRB~990123 host galaxy broad
         band rest-frame ($z=1.6$)
         spectrum to spectrum of S1-galaxies $\log F_{\lambda}=\log F_{\lambda, S1}+C$
	 (see Connoly et al.,~1995). 
	 FWHM of each filter for $\lambda_{eff}$ with the account for $z$
	 are denoted by dashed horizontal lines with bars.}
\label{grb990123_S1}
\end{center}
\end{figure*}

However, we consider it to be not quite correct to compare our
broad band spectra to normal Hubble types of galaxies. Obviously the
starburst activity may drasticaly change the spectral distribution
towards the ultraviolet part of the spectrum. According to this consideration,
we compared our $BVR_cI_c$ spectra to the averaged spectral energy distribution
of starburst galaxies from Connoly et al.~(1995).
Figure~\ref{grb980703_S2} and Figure~\ref{grb990123_S1} demostrate a
comparision of the starburst averaged 
spectral energy distributions to the $BVR_cI_c$ broad band spectra of the host
galaxies GRB~980703 and GRB~990123, respectively.
The spectra of sturburst were grouped according to increasing values of
the color excess $E(B-V)$: from S1, with $E(B-V)=0.05$ to S6, with
$E(B-V)=0.7$ (Connoly et al. 1995). Using relation for $\tau_B^l$
(Balmer optical depth) from
Calzetti et al. (1994) we derived the values of color excess for
individual starburst galaxies. It are $E(B-V)<0.10$ for S1, $0.11<E(B-V)<0.21$
for S2,  $0.25<E(B-V)<0.35$ for S3, $0.39<E(B-V)<0.50$ for S4,
$0.51<E(B-V)<0.60$ for S5 and $0.61<E(B-V)<0.70$ for S6 (see Table 3
in Calzetti et al. 1994).
In Figure~\ref{grb980703_S2} and Figure~\ref{grb990123_S1} the spectra of
the S1 and S2 type galaxies was averaged with a 10\AA\ window.

In this case, the calculations of the K-corrections from equation \ref{k-corr}
yield: $K_B = -0.01$ for the host galaxy of GRB~980703 and
       $K_B = 0.13$ for the host galaxy of GRB~990123.
Then, the absolute magnitudes for the host galaxy of GRB980703 are:
$M_{B_{rest}}=-20.60$ for model~(A),
$M_{B_{rest}}=-21.12$ for model~(B) and
$M_{B_{rest}}=-21.73$ for model~(C).
For the host of GRB990123 $M_B$ are:
$M_{B_{rest}}=-20.20$ for model~(A),
$M_{B_{rest}}=-21.02$ for model~(B) and
$M_{B_{rest}}=-21.82$ for model~(C).

\section{Estimations of star-forming rate}
We have roughly estimated also SFR using the continuum luminosity at
$\lambda_{rest}=2800\mbox{\AA}$ (see Madau et al.~1998).
In the calculations we assumed the cosmological models described above. \par
For the host galaxy of GRB~980703 the effective wavelength of the $V$ band
for $z=0.966$ corresponds to $2800\mbox{\AA}$ in the rest frame. Using the
flux in the $V$ band we estimated SFR for the host galaxy of GRB980703:
$~SFR_{Salpeter} \approx 15 \pm 2 \ M_\odot\ yr^{-1}$,
$~SFR_{Scalo} \approx 23 \pm 2 \ M_\odot\ yr^{-1}$ for model~(A);
$~SFR_{Salpeter} \approx 24 \pm 2 \ M_\odot\ yr^{-1}$,
$~SFR_{Scalo} \approx 37 \pm 4 \ M_\odot\ yr^{-1}$ for model~(B);
$~SFR_{Salpeter} \approx 43 \pm 4 \ M_\odot\ yr^{-1}$,
$~SFR_{Scalo} \approx 66 \pm 6 \ M_\odot\ yr^{-1}$ for model~(C),
where the index of Salpeter and Scalo denotes the Salpeter and Scalo
initial mass function (IMF) (Madau et al.~1998). \par
To estimate SFR of the host galaxy of GRB~990123, we used the interpolated
value of the flux at the wavelength $2800\mbox{\AA}$ in the rest frame
between $R_c$ and $I_c$ band. We assumed
$$\log F_{\nu, 2800\mbox{\AA}} =
-29.28 \pm 0.13 \ \frac{erg}{s\cdot cm^2 Hz}\ .
$$
The calculation yields:
$~SFR_{Salpeter} \approx 8 \pm 2 \ M_\odot\ yr^{-1}$,
$~SFR_{Scalo} \approx 12 \pm 4 \ M_\odot\ yr^{-1}$ for model~(A);
$~SFR_{Salpeter} \approx 17 \pm 6 \ M_\odot\ yr^{-1}$,
$~SFR_{Scalo} \approx 25 \pm 9 \ M_\odot\ yr^{-1}$ for model~(B);
$~SFR_{Salpeter} \approx 34 \pm 12 \ M_\odot\ yr^{-1}$,
$~SFR_{Scalo} \approx 54 \pm 17 \ M_\odot\ yr^{-1}$ for model~(C). \par
Of course, these stimates are the lower limit to SFR because our
calculations were performed without any galaxy rest-frame extinction
correction. Moreover, uncertaines of our results are estimated formally
from the errors of fluxes. 

\section{Discussion}
Observations were carried out a significant time
after the gamma-ray bursts. For the host galaxy of GRB~980703 it was about 20 days
after the gamma-ray burst, and for the GRB~990123 host galaxy about half a year.
This allows us to consider that  the contribution of the optical transient is
very small and we observed light only of the host galaxies. \par
As a discussion, it should be noted that there are uncertainties in
the estimates of the absolute magnitudes of the host galaxies due to the
K-correction. In the case of the host galaxy of GRB~990123 this uncertainty is
about $1^m$, and in the case of the host of GRB~980703 it is about $0^m.7$.
However, we consider that a more correct result is the value of $M_B$
according to the comparision to the starburst averaged spectra ---
$M_{Brest} = -20.60, -21.12, -21.73$ for the host galaxy of GRB~980703
in (A), (B), (C) cosmological models respectively, and
$M_{Brest} = -20.20, -21.02, -21.82$ for the host galaxy of GRB~990123.
Moreover,
our $BVR_cI_c$ broad band spectra are better fitted by the S1 and S2 spectral
energy distribution.
To compare our results to the results of Bloom et al. (1998,1999) we assume a
cosmology
model with $H_0 = 65 \ km\ s^{-1}\ Mpc^{-1}$, $\Omega_m = 0.2$ and
$\Omega_\Lambda = 0$. For the GRB~980703 host galaxy we used the value of
the luminosity distance from Bloom et al. (1998),
$d_L = 1.92 \times 10^{28} cm$.
In the case of GRB~990123 host galaxy we used the value of the luminosity
distance from Bloom et al. (1999), $d_L = 3.7 \times 10^{28} cm$.
Calculations from equation \ref{absmagn} yield:
$M_{B_{rest}} = -20.8$ and $M_{B_{rest}} = -20.62$ for the GRB~980703 and
GRB~990123 host galaxies respectively, while the values of Bloom et al.
are $-20.2$ and $-20.0$ for the GRB~980703 and GRB~990123 host galaxies,
respectively.
Note that the estimates of absolute magnitudes of
Bloom et al. (1998, 1999) for the both host galaxies are performed
in another way without the K-correction by fitting to the spectrum in the case
of the GRB~980703 host galaxy and by interpolating between the observed
the STIS and the $K$-band data points
using a power law in the case of GRB~990123 host galaxy. \par
It should be noted that our estimates of the star-forming rate are higher
than the estimates of Djorgovski et al.~(1998) and Bloom et al.~(1999).
It is interesting that the
for GRB~980703 host galaxy our flux at $\lambda=2800\mbox{\AA}$ is
$\sim 3.1 \mu Jy$ and is matching the value of the flux on July 7 from
Djorgovski et al.~(1998).
Probably, the OT contribution was already negligible in the $V$ band on
July 7.
In the case of GRB~990123 host galaxy our value of the flux at
$\lambda=2800\mbox{\AA}$ was estimated by interpolation between observed
points. However, the values of Bloom et al. (1999) are $0.17 \mu Jy$
(for $\beta=0$) and $0.21 \mu Jy$ (for $\beta=-0.8$) while our ones are
$0.52_{-0.14}^{+0.18} \mu Jy$. This discrepancy can be explain as follow.
Our estimate of the flux was performed by interpolation between the $R_c$
($\lambda_{eff}=6588\mbox{\AA}$) and $I_c$ ($\lambda_{eff}=8060\mbox{\AA}$)
bands, while the values of Bloom et al. were interpolated with power law
between STIS (approximately $V$ band, $\lambda_{eff}=5505\mbox{\AA}$) point
and $K$ band ($\lambda_{eff}=2.195\mu m$). Moreover, Bloom et al. measured
the flux of the host galaxy by masking sets of pixel dominated by the OT
(Bloom et al. 1999) because observations in the $K$ band were carried out
on epoch 9 and 10 Febrary UT, 17-18 days after gamma-ray burst,
(Bloom et al. 1999) when contribution of the OT
was not negligible, while our observations was performed about a half year
after gamma-ray burst occured. Obviously, the our estimate is more exact
than that of Bloom et al.

\acknowledgements{ We thank S.V. Zharikov for assistance in observations,
and T.N. Sokolova for the help in work with the text.}
 This work has been partly supported by the "Astronomy" Foundation
 (grant 97/1.2.6.4), INTAS N 96-0315 and RFBR N98-02-16542.

\end{document}